\documentclass[a4paper,14pt]{extarticle}

\usepackage{cmap}
\usepackage[cp1251]{inputenc}
\usepackage[english]{babel}
\usepackage{braket,amssymb,amsmath,array,commath,mathtext,wrapfig,tikz}
\usepackage[arrows=pgf-filled,version=3]{mhchem}
\usepackage{bm}
\usepackage{amsthm,amsfonts,amscd}
\usepackage[super,compress]{cite}
\usepackage{graphicx}
\usepackage{lscape}
\usepackage{geometry}
\usepackage{indentfirst}
\usepackage{afterpage,soul,ulem,dcolumn,changebar,longtable,hhline,multirow,makecell}
\usepackage[small,bf]{caption}
\begin{document}
	
	\begin{center}
		\textbf{The structure of the generalized Vaidya spacetime containing the eternal naked singularity}
	\end{center}
	
	\begin{center}
		\textbf{Vitalii Vertogradov}
	\end{center}
	
	\begin{center}
		Physics department, Herzen state Pedagogical University of Russia,
		
		48 Moika Emb., Saint Petersburg 191186, Russia
		
		SPb branch of SAO RAS, 65 Pulkovskoe Rd, Saint Petersburg 196140, Russia
		
		vdvertogradov@gmail.com
	\end{center}
	
	\section{Abstract} In this paper the structure of the generalized Vaidya spacetime when the type-II of the matter field satisfies the equation of the state $P=\rho$ is investigated. Satisfying all energy conditions we show that this spacetime contains the 'eternal' naked singularity. It means that once the singularity is formed it will never be covered with the apparent horizon. However, in the case of the apparent horizon formation the resulting object is a white hole. We also prove that this spacetime contains only null naked singularity.
	
	\textit{Generalized Vaidya spacetime; Naked singularity; White Hole; Vectons}
	
	\section{Introduction}
	
	At the end of its lifetime cycle the massive star undergoes the continuous gravitational collapse. Oppenheimer and Snider were the first who were able to construct the model of gravitational collapse of the dust cloud~\cite{bib:open}. The result of this collapse is a black hole and it was believed that the result of such a process is inevitably a black hole. There is a singularity at the centre of the spherically symmetric black hole. According to the cosmic censorship conjecture [CCC] any singularity must be hidden under the event horizon. However, Papapetroo~\cite{bib:pap} was one of the first who showed that Vaidya spacetime ~\cite{bib:vay}contains the naked singularity. So Vaidya spacetime is one of the earliest counterexamples of CCC violation. After that it was understood that the naked singularity might be the result of the continuous gravitational collapse~\cite{bib:joshi} (For more detail information see \cite{bib:joshi2}).
	
	When we speak about the result of gravitational collapse, then we should realize that it might be not only a black hole but also the naked singularity. The result of such a process depends upon the initial data. One should understand that this nakedness is temporary and within the short period of time the singularity will be covered by the apparent horizon. Hence there is a question about the existence of the eternal naked singularities. Under the notion 'eternal' we assume that the singularity was formed, but will never be covered with the apparent horizon. It was shown that in the case of the gravitational collapse of the scalar field~\cite{bib:joshi3} and the generalized Vaidya spacetime~\cite{bib:ver1} the result might be the eternal naked singularity formation. However, in these models one has to violate the strong energy condition~\cite{bib:pois} and assumes that the matter possesses the negative pressure. So there is a question if we can construct the model of the eternal naked singularity formation without violating any energy condition.
	
	In this paper, we consider the generalized Vaidya spacetime~\cite{bib:vunk} when the type II of the matter field satisfies the equation of the state $P=\rho$, where $P$ and $\rho$ are pressure and energy density respectively. The type-I of the matter field represents the null dust. The equation of the state $P=\rho$ is so-called the extremely rigid equation of the state and it corresponds to the particle - vecton~\cite{bib:nov}. We can also notice that the extremely rigid equation of the state corresponds to the case when the speed of the sound is equal to the speed of light and often the matter is called as the stiff fluid. In this article we will show that this space-time contains the eternal naked singularity and also a white hole.
	The paper is organized as follows: in sec. II we consider the generalized Vaidya spacetime when the II-type of the matter spacetime possesses the extremely rigid equation of the state. In sec. III we consider the eternal naked singularity formation. In sec. IV white hole formation is investigated. Sec. V is the conclusion.
	Throughout the paper $c=G=1$ system of units will be used.
	
	\section{Generalized Vaidya spacetime \\ with the extrimely rigid equation of the state}
	
	The generalized Vaidya spacetime has the following form~\cite{bib:vunk}:
	\begin{equation}
		\begin{split}
			\label{eq:met}
			ds^2=-\left ( 1-\frac{2M(r,v)}{r} \right ) dv^2+2\varepsilon dvdr+r^2 d\Omega^2 \,, \\
			d\Omega^2=d\theta^2+\sin^2\theta d\varphi^2 \,,
		\end{split}
	\end{equation}
	here $M(v,r)$ - the mass function depending on coordinates $r$ and $v$ which corresponds to advanced/retarded time, $\varepsilon=\pm 1$ depending on ingoing/outgoing shells.
	We can write down the energy momentum tensor in the following form:
	\begin{equation}
		T_{ik}=T^{(n)}_{ik}+T^{(m)}_{ik}\,,
	\end{equation}
	where the term $T^{(m)}_{ik}$ corresponds to the matter field of the type-I and the other $T_{ik}^{(n)}$ corresponds to the matter field of the type-II~\cite{bib:hok}.
	Now let us write down the expression of the energy momentum tensor:~\cite{bib:vunk}
	\begin{equation} \label{eq:ten}
		\begin{split}
			T^{(m)}_{ik}= \mu L_{i}L_{k}\,, \\
			T^{(n)}_{ik}=(\rho+P)(L_{i}N_{k}+L_{k}N_{i})+Pg_{ik} \,, \\
			\mu=\frac{2 \varepsilon \dot{M}}{ r^2} \,, \\
			\rho=\frac{2M'}{r^2} \,, \\
			P=-\frac{M''}{r} \,, \\
			L_{i}=\delta^0_{i} \,, \\
			N_{i}=\frac{1}{2} \left (1-\frac{2M}{r} \right )\delta^0_{i}-\varepsilon \delta^1_{i} \,, \\
			L_{i}L^{i}=N_{i}N^{i}=0 \,, \\
			L_{i}N^{i}=-1 \,.
		\end{split}
	\end{equation}
	here $P$ - pressure, $\rho$ - density, $\mu$ - the energy density of the null dust. And $L,N$ - two null vectors.
	This model must be physically reasonable, so the energy momentum tensor should satisfy weak, strong and dominant energy conditions. It means that $\rho$ must be positive and for any non-spacelike vector $v^{i}$:
	\begin{equation} \label{eq:emt}
		T_{ik} v^{i} v^{k}>0\,,
	\end{equation}
	and the vector $T_{ik}v^{i}$ must be timelike.
	Strong and weak energy conditions demand:
	\begin{equation}
		\begin{split} \label{eq:ws}
			\mu \geq 0 \,, \\
			\rho \geq 0 \,, \\
			P\geq 0 \,.
		\end{split}
	\end{equation}
	The dominant energy condition imposes following conditions on the energy momentum tensor:
	\begin{equation} \label{eq:dom}
		\begin{split}
			\mu \geq 0 \,, \\
			\rho \geq P\geq 0\,.
		\end{split}
	\end{equation}
	The properties of the generalized Vaidya spacetime have been studied for the equation of the state $P=\alpha \rho$ where $\alpha$ belongs to the interval $(-1, \frac{1}{3}$ in articles~\cite{bib:ver3, bib:ver2, bib:ver4}. If we satisfy this equation of the state, then the mass function $M(r,v)$ has the form
	\begin{equation} \label{eq:mass}
		M(r,v)=C(v)+D(v)r^{1-2\alpha}\,,
	\end{equation}
	where $C(v)$ and $D(v)$ are positive functions of time $v $.
	One should note that the speed of sound in the medium depends upon $\alpha$ in the following way :
	\begin{equation} \label{eq:sound}
		v_{sound}=\sqrt{\alpha} \,.
	\end{equation}
	We want to study the generalized Vaidya spacetime with the extremely rigid equation of the state. For this purpose, we must consider the equation of the state:
	\begin{equation}
		P=\rho \,.
	\end{equation}
	In this case the mass function has the form \eqref{eq:mass} with $\alpha=1$. One can see from \eqref{eq:sound} that the speed of sound in this case is equal to the speed of light. However, we must consider function $D(v)$ properly. Now it can't be positive because the density expression is given by:
	\begin{equation}
		\rho= -\frac{D(v)}{r^4} < 0 \,,
	\end{equation}
	and if $D(v)>0$ then we violate energy conditions \eqref{eq:ws} and \eqref{eq:dom}. So to satisfy these conditions we must assume that $D(v)<0$ or we can write:
	\begin{equation} \label{eq:neg1}
		D(v)=-D'(v) \,,
	\end{equation}
	where now function $D'(v)$ is positive.
	We want to investigate the question about the singularities in this spacetime which are formed at $v=0$ and $r=0$. But in this case we must impose some extra conditions on functions $C(V)$ and $D'(v)$ not to violate energy conditions \eqref{eq:ws} and \eqref{eq:dom} because if $\dot{C}(v)<\frac{\dot{D'}(v)}{r}$ at some point then $\mu$ becomes negative and we violate our energy conditions. So we must consider two cases:
	\begin{enumerate}
		\item $D'(v)\equiv \lambda$, where $\lambda$ is positive real constant. In this case $\dot{D'}(v)\equiv 0$ and if we impose the following condition $C(V) \geq 0$ then we satisfy all energy condition.
		\item $D'(0)=0$ but in this case even if $\lim\limits_{v\to 0, r\to 0} \frac{D'(v)}{r}=X_1$ where $X_1$ finite positive constant and if $C(v)>X_1$ then anyway we would violate energy conditions at the later stage.
	\end{enumerate}
	
	Also $C(v)$ must satisfy the following condition:
	\begin{equation}
		\begin{split}
			\dot{C}(v)\geq 0\, \mathrm{if} \, \varepsilon=+1 \,, \\
			\dot{C}(v)\leq 0 \, \mathrm{if} \, \varepsilon=-1 \,.
		\end{split}
	\end{equation}

	\section{The structure of the singularity }
	The apparent horizon equation in the case of generalized Vaidya spacetime with extremely rigid equation of the state is given by:
	\begin{equation}
		\frac{2C(v)}{r}-\frac{2\mu}{r^2}-1=0 \,.
	\end{equation}
	The time of the singularity formation is $v=0$. So solving the above equation, we obtain:
	\begin{equation} \label{eq:vad}
		\begin{split}
			R^2-2C(v)r+2\mu =0 \,, \\
			\frac{D}{2}=C^2(v)-2\mu \,.
		\end{split}
	\end{equation}
	From this equation we can conclude that if
	\begin{equation}
		C^2(v)< 2\mu \,,
	\end{equation}
	then the apparent horizon will be never formed.
	Now let's consider the question about the existence of null radial geodesics which terminate at the central singularity in the past. The radial null geodesic equation has the following form:
	\begin{equation} \label{eq:geo}
		\frac{dv}{dr}=\frac{2\varepsilon r^2}{r^2-2C(v)r+2\mu} \,.
	\end{equation}
	If we consider the limit $\lim\limits_{v\to 0, r\to 0} \frac{dv}{dr}$ then we can see that In the denominator of this fraction \eqref{eq:geo} we have a finite positive number (see \eqref{eq:vad}.). And as the result we have:
	\begin{equation}
		\begin{split}
			X_0=\lim\limits_{v\to 0, r\to 0} \frac{dv}{dr}=0 \,, \\
			v=\text{const}\,.
		\end{split}
	\end{equation}
	Usual and generalized Vaidya spacetimes we obtain from Schwarzschild spacetime assuming that the time axe is the radial null geodesic in Schwarzschild spacetime because of it, we can have $X_0=0$ as the condition for the existence of the radial null geodesic. It is also because to obtain the Vaidya metric one should do the following coordinate transformation of the Schwarzschild spacetime:
	\begin{equation}
		v=t+\varepsilon (r+2m\ln \left |r-2m \right | ) \,.
	\end{equation}
	So in the case $v=const.$ we have the eternal naked singularity formation\footnote{Under the notion 'eternal' we understand that the singularity is formed and will be never covered with the apparent horizon}. However, not not all $v=const.$ suits us, the result depends on $\varepsilon$. If $\varepsilon=+1$ - falling matter, then the null adial geodesic $v=const.$ is past oriented. So to have future oriented radial null geodesic $v=const.$ which originate at the central singularity in the past one should consider outgoing Vaidya spacetime $\varepsilon=-1$ i.e. the radiating Vaidya spacetime.
		\subsection{The radial timelike geodesics}
		We have found out that there is a family of the radial null future-directed geodesics which terminate at the central singularity in the past. Now we are interested if not only massless particles can move away from the central singularity but also particles which possess the mass. For this purpose, we must consider the timelike geodesic equation. We can notice that the generalized Vaidya spacetime extremely rigid equation of the state doesn't depend on the angle $\varphi$. Hence we can conclude that the angular momentum $L$ is conserved. However, the metric depends on the time $v$ so the energy is not constant, but some function of the time $v$.
		To derive the timelike geodesic equation, we should consider the lagrangian. In this case it has the following form:
		\begin{equation} \label{eq:lag}
			2\mathcal{L}=-\left (1-\frac{2C(v)-\frac{2\mu}{r}}{r} \right ) \dot{v}^2+2\dot{v}\dot{r}+r^2(\dot{\theta}^2+\sin^2\theta \dot{\varphi}) \,,
		\end{equation}
		where sign dot denotes the partial derivative on affine parameter $\tau$. We are interested in the radial geodesics so because of it, we can demand $\dot{\theta}=\dot{\varphi}=0$. Taking into account this one and \eqref{eq:lag} we can find the energy $E$ as a function of the time $v$:
		\begin{equation} \label{eq:ene}
			-E(v)=\frac{\partial \mathcal{L}}{\partial \dot{v}}=-\left (1-\frac{2C(v)-\frac{2\mu}{r}}{r} \right )\dot{v}+\dot{r} \,.
		\end{equation}
		We impose the condition $g_{ik}\dot{x}^i\dot{x}^k=-1$ to obtain the timelike geodesic. In our case this condition gives:
		\begin{equation} \label{eq:rad}
			-1=(-E+\dot{r})\dot{v} \,.
		\end{equation}
		Now if we use \eqref{eq:ene} the equation \eqref{eq:rad} gives:
		\begin{equation} \label{eq:1}
			\dot{r}=\pm \sqrt{E^2(v)-\left (1-\frac{2C(v)-\frac{2\mu}{r}}{r} \right )} \,.
		\end{equation}
		Now substituting \eqref{eq:1} into \eqref{eq:ene} we obtain:
		\begin{equation} \label{eq:2}
			\dot{v}=\frac{-E- \sqrt{E^2(v)-\left (1-\frac{2C(v)-\frac{2\mu}{r}}{r} \right ) }}{-\left (1-\frac{2C(v)-\frac{2\mu}{r}}{r} \right )} \,,
		\end{equation}
		we picked up sign ''+'' because we are interested in outgoing geodesics.
		Now combining \eqref{eq:1} and \eqref{eq:2} we finally obtain the timelike geodesic equation:
		\begin{equation}\label{eq:tim}
			\frac{dv}{dr}=\frac{-E-\sqrt{E^2(v)-\left (1-\frac{2C(v)-\frac{2\mu}{r}}{r} \right ) }}{-\left (1-\frac{2C(v)-\frac{2\mu}{r}}{r} \right )\sqrt{E^2(v)-\left (1-\frac{2C(v)-\frac{2\mu}{r}}{r} \right ) }} \,.
		\end{equation}
		Now if we consider the limit $\lim\limits_{v\to 0, r\to 0} \frac{dv}{dr}$ then we can easily see that this limit is equal to zero. Hence we can conclude that the equation \eqref{eq:tim} has the solution $v=\text{const.}$. But the case $v=\text{const.}$ is suited only for the radial null geodesics (not for timelike) so we can conclude that there exists only the radial null geodesic which terminates at the central singularity int the past.
		\section{The white hole formation}
		We can see that \eqref{eq:vad} can have positive solutions if
		\begin{equation}
			C^2(v) \geq 2\mu \,.
		\end{equation}
		So we have solutions:
		\begin{equation}
			r_{\pm}=C(v)\pm \sqrt{C^2(v)-2\mu} \,.
		\end{equation}
		It is worth notecing that these roots are both positive. Let's consider the region $0\leq r \leq r_-$. The existence of a famile of the radial null geodesics which terminate at the central singularity in the past we proved in the previous section. Also, we can notice that the limit $\lim\limits_{r\to 0, v\to 0} \frac{dv}{dr} =0$ i.e. this limit is not negative and it means that we have the movement only from the central singularity and we don't have the movement towards it. Now we should prove that the expansion $\theta$~\cite{bib:pois} is positive everywhere in this region. The expansion in the case of generalized Vaidya spacetime extremely rigid equation of the state has the following form~\cite{bib:ver1}:
		\begin{equation}
			\theta=e^{-\gamma}\frac{2}{r^3}\left (r^2-2C(v)r+2\mu \right ) \,,
		\end{equation}
		where the parameter $\gamma$ doesn't have any impact on the sign $\theta$. So only the expression in the parentheses has the impact of the sign $\theta$. But from \eqref{eq:vad} we can notice that this expression is positive in the region $0\leq r \leq r_-$.
		We have proved that the expansion $\theta$ is positive in the region $0\leq r \leq r_-$. Hence this region corresponds the white hole solution.
		\section{Conclussion}
		In this article we considered the particular solution of the generalized Vaidya spacetime when the type-II of the matter fields satisfies the extremely rigid equation of the state$P=\rho$. This solution corresponds to vecton particles. In this spacetime the speed of sound is equal to the speed of light. In this case, satisfying all energy conditions, we prove that this spacetime contains the 'eternal' null naked singularity. However, this nakedness is possible only for radiating metric, i.e. $\varepsilon=-1$. The other possibility is the white hole formation. In comparison with naked singularity the white hole also contains the family of non-spacelike future-directed geodesics which terminates at the central singularity in the past. But as compared to the naked singularity, the white hole singularity is covered with apparent horizon.
		
		One should note that the case of the stiff fluid is similar to the usual charged Vaidya solution \footnote{Under 'usual' we understand that the mass function depends only on time $v$ but the black hole possesses the charge $Q$.}. This metric doesn't violate the weak energy condition~\cite{bib:vay2} but one can maintain this condition due to charge and the Lorentz force. However, here the function $D(v)\neq Q(v)$ because in the case of the absence of the matter field type-II $D(v)\equiv 0$. Thus, there is no any Lorentz force which can prevent the violation of the weak energy condition in the general case because of it we have to demand the restrictions \eqref{eq:neg1}.
		
		We constructed the model of the 'eternal' naked singularity without violating any energy conditions. However, this singularity is not timelike. To obtain the timelike 'eternal' naked singularity we either should consider another the equation of the state or another spacetime. Obtaining of timilike eternal naked singularity is the question of future investigations.
		
		\textbf{acknowledgments} The author says thanks to professor Pankaj Joshi for scientific discussion and grant Num. 22-22-00112 RSF for financial support.

	\end{document}